\ifx\pdfoutput\undefined        
  \documentclass[twocolumn]{article}
  \usepackage{url}
\else                           
  \documentclass[pdflatex,twocolumn]{article}
  \usepackage[bookmarks=false]{hyperref}
\fi

\usepackage{graphicx}
\usepackage{url}
\graphicspath{{./fig/}{./png/}}

\newcommand{\EQ}{\begin{equation}}
\newcommand{\EN}{\end{equation}}
\newcommand{\EQA}{\begin{eqnarray}}
\newcommand{\ENA}{\end{eqnarray}}

\newcommand{\Fig}[1]{Fig.~\ref{#1}}

\newcommand{\bra}[1]{\langle #1\rangle}

\newcommand{\dd}{{\rm d} {}}

\newcommand{\ypnas}[5]{ (#1) #5, {\em Proc.\ Natl.\ Acad.\ Sci.\ }{\bf #2}, #3--#4.}

\newcommand{\ysci}[5]{ (#1) #5, {\em Science }{\bf #2}, #3--#4.}

\newcommand{\ynat}[5]{ (#1) #5, {\em Nature }{\bf #2}, #3--#4.}

\newcommand{\yphl}[5]{ (#1) #5, {\em Phys.\ Lett.\ } {\bf #2}, #3--#4.}
\newcommand{\yoleb}[5]{ (#1) #5, {\em Orig.\ Life Evol.\ Biosph.\ }{\bf #2}, #3--#4.}
\newcommand{\yab}[5]{ (#1) #5, {\em Astrobiol.\ }{\bf #2}, #3--#4.}
\newcommand{\yjour}[6]{ (#1) #6, {\em #2} {\bf #3}, #4--#5.}

\newcommand{\ppjour}[4]{ (#1) #4, {\em #2} {\bf #3} (in press).}

\newcommand{\poleb}[2]{ (#1) #2, {\em Orig.\ Life Evol.\ Biosph.} (in press).}

\begin{document}

\title{Spatial dynamics of homochiralization}
\author{Tuomas Multam\"aki and Axel Brandenburg\\
NORDITA, Blegdamsvej 17, DK-2100 Copenhagen \O, Denmark
}

\date{
Received March 15, 2005;
accepted May 4, 2005}

\maketitle

\abstract{
The emergence and spreading of chirality
on the early Earth is considered by studying a set of 
reaction-diffusion equations based on a polymerization model.
It is found that effective mixing of the early oceans
is necessary to reach the present homochiral state.
The possibility of introducing mass extinctions and
modifying the emergence rate of life is discussed.
}

\section*{Introduction}

The question of the origin of life is closely connected with the question
of how handedness (chirality) of living organisms came about
(Avetisov et al.\ 1991).
The sugars produced in plants via photosynthesis, for example via the
rather innocently looking reaction
\EQ
6\,\mbox{CO}_2+6\,\mbox{H}_2\mbox{O}\rightarrow
\mbox{C}_6\mbox{H}_{12}\mbox{O}_6+6\,\mbox{O}_2,
\EN
are all right-handed and rotate the polarization plane of polarized light
in a right-handed sense, i.e.\ the sugar is said to be dextrorotatory.
This preferred handedness is remarkable because the difference in
binding energy between right-handed and
left-handed (levorotatory) sugars is extremely tiny
(relative energy difference: $10^{-17}$).
Once dead, the preferred handedness gradually disappears until there
are equally many right and left handed molecules.
This process, which speeds up with increasing temperature, is called
racemization and can even be used as an approximate dating method
(Hare \& Mitterer 1967, Bada et al.\ 1970).
For a simpler and closely related reaction that displays a similar
behavior we refer to the formose reaction discussed by Toxvaerd (2005).

Another type of dextrorotatory sugar with five carbon atoms occurs in the
backbone of RNA, for example, making RNA therefore chiral.
RNA can have autocatalytic properties, which led to the idea of an early
``RNA world'' (Gilbert 1986, Joyce 1991) where naked RNA molecules
could have catalyzed the polymerization of other RNA molecules.
But since the RNA molecule consists of right-handed sugars, chirality
selection must have occurred at an even earlier stage.

Two possibilities are commonly discussed.
Either chirality was imposed by physical factors such as magnetic
fields, polarized light from a nearby pulsar, or even the tiny energetic
preference due to the parity-breaking electroweak interaction.
In such a scenario, homochirality may well have been a
{\it pre\-re\-qui\-site} for the origin of life, allowing for example for
the assembly of structurally more stable polymers.
The difficulty here is that such mechanisms would hardly explain the
complete homochirality observed in living matter.
The other possibility is that homochirality was a {\it consequence} of
life, and that the presently prevailing handedness was selected
spontaneously during the assembly of the first polymers.
As shown in the early paper by Frank (1953), such a process requires the
production of left and right handed molecules to be mutually antagonistic.
He devised pairs of ordinary differential equations demonstrating how
this could possibly be described mathematically.

The latter scenario is the one we consider also in the present paper.
Significant progress has been made since Frank's early paper.
In particular, the origin of the mutually antagonistic behavior has been
identified to be the so-called enantiomeric cross-inhibition, a process
that spoils further polymerization once a monomer of opposite handedness
has been attached to an already existing polymer.
The evidence for the occurrence of this process is entirely experimental
and goes back to the early work of Joyce et al.\ (1984), who found that
in non-enzymatic template-directed polymerization of RNA strands,
only a homochiral supply of mononucleotides that are complementary
to the template can polymerize to a typical length of 20 nucleotides.
Even a small amount of mononucleotides of opposite chirality prevents the
formation of longer polymers, as is seen in high performance color
chromatograms.
Similar experiments have subsequently also been carried out by
Schmidt et al.\ (1997) and Kozlov et al.\ (1998), for example.
In the following we describe in more detail how this process can be modeled.

\section*{The polymerization model}

In the polymerization model of Sandars (2003) large polymers 
are generated by joining successive monomers into long chains.
Monomers can be both left, $L_1$, and right, $R_1$, handed, and longer chains
are formed according to the following set of reaction equations:
\begin{eqnarray}
L_{n}+L_1&\stackrel{2k_S~}{\longrightarrow}&L_{n+1},
\label{react1}\\
L_{n}+R_1&\stackrel{2k_I~}{\longrightarrow}&L_nR_1,
\label{LnR1react}
\label{react2}\\
L_1+L_{n}R_1&\stackrel{k_S~}{\longrightarrow}&L_{n+1}R_1,
\label{react3}\\
R_1+L_{n}R_1&\stackrel{k_I~}{\longrightarrow}&R_1L_nR_1,
\label{react4}
\end{eqnarray}
where $k_S$ and $k_I$ are the reaction coefficients for monomers
to a polymer of the same or of the opposite handedness, respectively.
For all four equations we also have the complementary reactions
obtained by exchanging $L\rightleftharpoons R$.
Note that chains are `spoiled' if a monomer of opposite chirality is
attached to the end of a longer chain, in which case the chain can only
grow at the other end. This is in essence what is meant by
enantiomeric cross-inhibition. As a results, polymers
such as $R_1L_2R_1$ and $L_1R_1$ can no longer grow.
As discussed in the introduction, this rule is motivated by the experiments
of Joyce et al.\ (1984) and others.

The monomers are initially generated from a substrate $S$ according to
\EQ
S\stackrel{k_C C_R~}{\longrightarrow} R_1,\quad
S\stackrel{k_C C_L~}{\longrightarrow} L_1,\quad
\EN
where $k_C$ is proportional to the regeneration rate of monomers, and
$C_R$ and $C_L$ determine the enzymatic enhancement of right and
left handed monomers. The dependence of $C_{L,R}$ on the existing
amount of polymer chains is essential to chirality selection as then
an existing excess of either chirality can be amplified. The
exact form of $C$ is not crucial and here we follow the choice 
$C_A=\sum n\, A_n,\ A=L,R$ (Brandenburg et al.\ 2005,
hereafter referred to as BAHN).
For alternative prescriptions of $C_A$ we refer to the papers by
Sandars (2003) and Wattis \& Coveney (2005).
The substrate itself is being replenished by a constant source term $Q$.

\begin{figure}[t!]
\begin{centering}
\includegraphics[width=5cm]{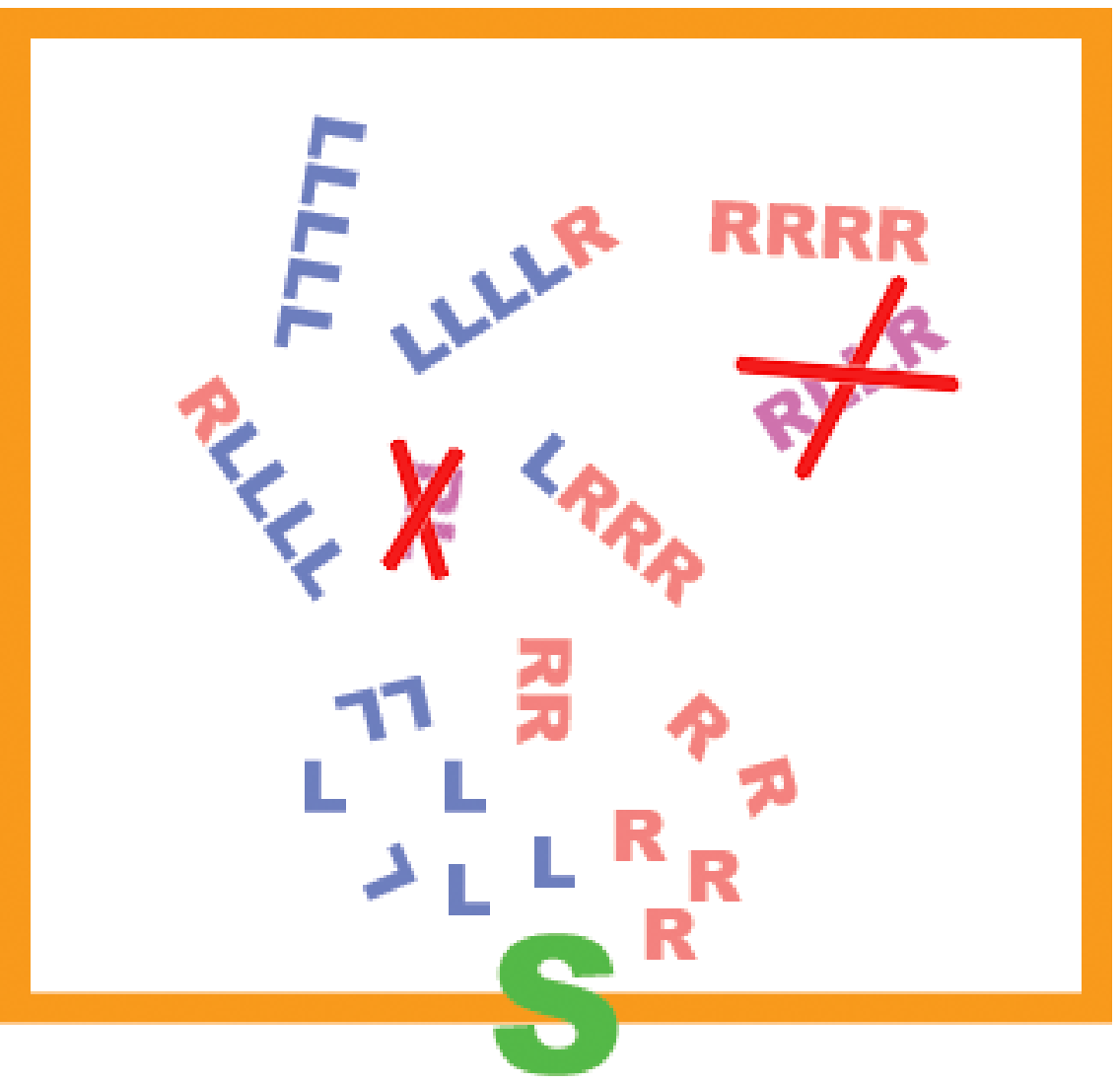}\caption{Polymerization process.
In addition to a number of monomers, L and R, there are several
isotactic dimers, LL and RR, as well as longer polymers.
Semi-spoiled polymers such as LLLLR and LRRR can still polymerize
on the unspoiled end.
Polymers such as LR and RLLR are dead and cannot polymerize further.}
\label{un1}
\end{centering}
\end{figure}

The polymerization process is represented pictorially in Fig.~\ref{un1},
where one can see how monomers begin to grow into longer chains and
can then be contaminated by a monomer of opposing chirality. The 
crossed out chains represent polymers that can no longer grow.

The feedback mechanism built into the polymerization model
leads to an unstable system that, depending on the fidelity $f$
of the enzymatic reactions, when perturbed from the initial racemic state 
can reach a homochiral state.
The prescription for $C_A$ discussed above assumes $f=1$.
If $f<1$, there is some `cross-talk' between $L$ and $R$ where $f$
determines the relative mixing between the two chiralities.

The chirality of a particular state is 
conveniently parameterized by a
parameter called the {\it enantiomeric excess}:
\EQ
\eta\equiv \frac{E_R-E_L}{E_R+E_L},
\EN
where $E_A=\sum n\, A_n$ with $A=L,R$. For $f$ larger than a critical value
(see BAHN for details), the racemic state ($\eta=0$) is unstable
with respect to small perturbations. Only for $f=1$, the end state
is fully homochiral ($\eta=\pm 1$).

\section*{Homochiralization in space}

The homochiralization process considered here and, indeed, in much of
the theoretical literature on the subject (e.g., Frank 1953, Wei-Min 1982,
Goldanskii and Kuzmin 1989, Avetisov \& Goldanskii 1993, Sandars 2003,
Saito and Hyuga 2004a) has been uniform in space.
Relaxing this restriction can, in principle, lead to completely new
chirality selection mechanisms that have no analogue in models without
spatial extend.
An example is the mechanism identified by Toxvaerd (2004) using
molecular dynamics simulations.
In the present work we simply extend the homochiralization process
discussed in the previous section to allow for the interaction with
neighboring regions through diffusion and/or advection (such as ocean
currents).
We begin with a state that is homochiral everywhere, but that there
are minute imbalances in space, i.e.\ infinitesimally small perturbations.
These perturbations grow locally, leading to patches with enantiomeric
excess of either handedness.
It is then natural to ask how a localized homochiral state spreads
to a racemic surroundings and, moreover, can regions of different
chirality coexist? This question is not restricted to the model
considered here and needs to be addressed in all models where
chirality is selected locally by a random process.

\begin{figure}[t!]
\begin{centering}
\includegraphics[width=5cm]{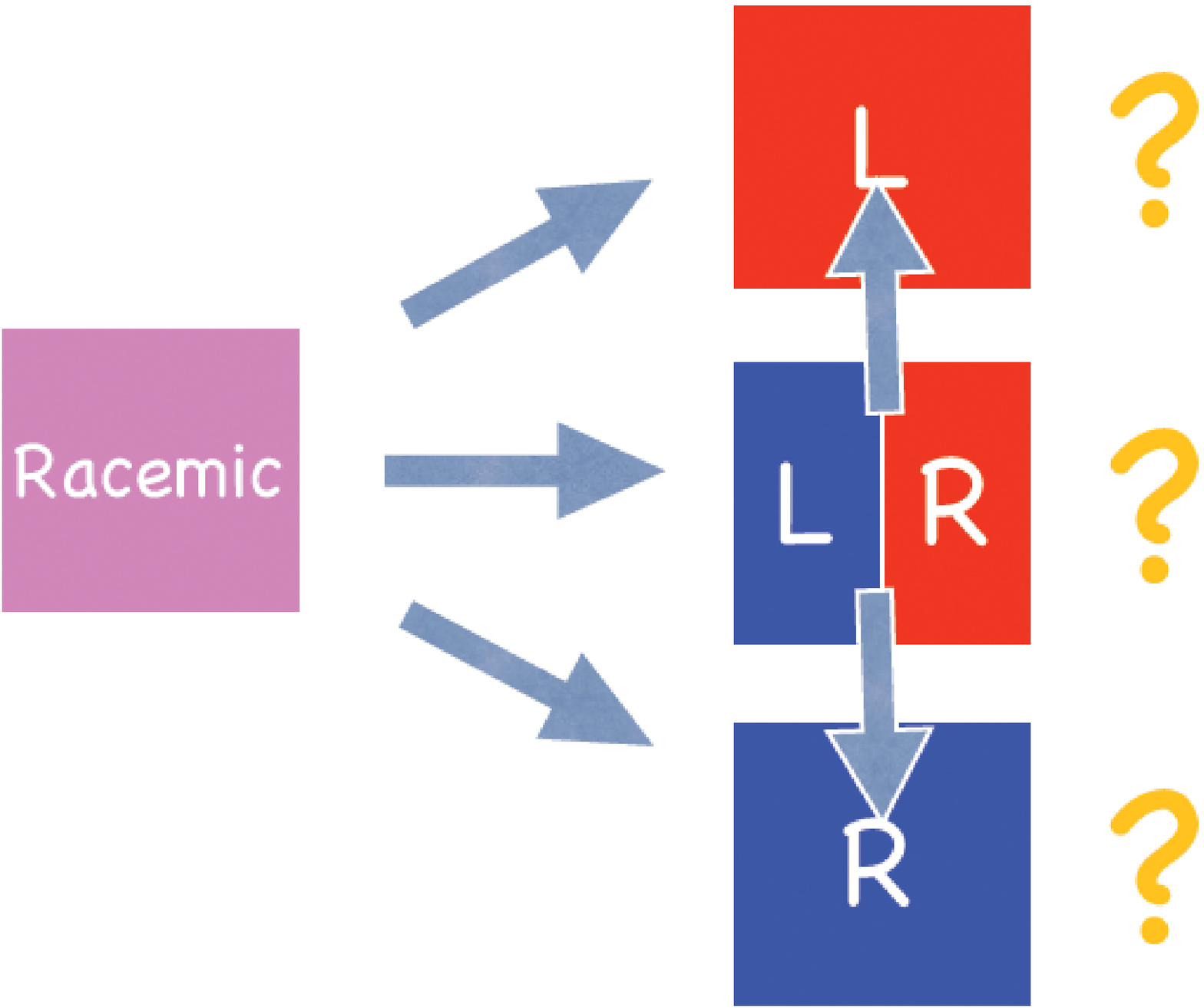}\caption{Different ways of reaching a 
homochiral state from an initial racemic state}
\label{un2}
\end{centering}
\end{figure}

The critical ingredients necessary to answer these questions are the
presence of different relevant
time scales. If the appearance of a homochiral region is a rare process,
say with a period of $\tau_{\rm life}$, and the global homochiralization
time scale, $\tau_{\rm global}$, is short compared to that, i.e.\
$\tau_{\rm global}\ll \tau_{\rm life}$, then one would not expect 
coexisting regions of 
different chirality.
In other words, if the emergence of life is a rare event, then
life forms with different handedness did not probably coexist.
This is the case modeled by Frank (1953) and many others after him.
In the other extreme case, homochiral regions appear frequently
compared to the speed at which they can dominate the early Earth, leading
to coexisting regions of opposite chirality (see \Fig{un2}). 
Even though here we are
mainly concerned with chirality, such arguments can be applied more
generally to the emergence and spreading of life as well.
For a discussion of the possibility of finding a second sample of
life on Earth see the recent paper by Davies \& Lineweaver (2005).

The spreading of chirality and coexistence of regions of different
chirality was studied in Brandenburg and Multam\"aki (2004),
hereafter referred to as BM.
There a reduced polymerization model with spatial extent was 
considered analytically and by numerical simulations.
The importance of spatial extent has already been emphasized by
Saito and Hyuga (2004b) who generalized the model of Saito and Hyuga (2004a)
by using a Monte-Carlo method as it is used in percolation studies.

\begin{figure}[t!]
\resizebox{\hsize}{!}{\includegraphics{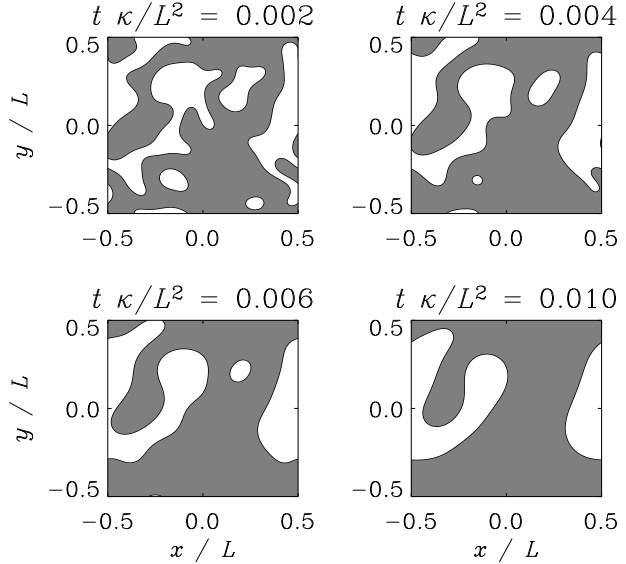}}\caption{
The spreading of homochiral regions in $2+1$ dimensions. The dark (light)
regions correspond to left (right) handed regions.
Time is given in global diffusion times, $\tau_{\rm turb}=L^2/\kappa$.
(Adapted from BM.)
}\label{pxxx}\end{figure}

The behavior
of the reduced model can be understood qualitatively: an initial
racemic mixture with small local perturbations quickly
relaxes locally to regions of different chirality. Hence, in this
work the emergence of life was considered to be a frequent and
rapid event. The oppositely handed regions then begin to spread
spatially into any possibly remaining racemic regions by
front propagation until they come in contact with a region of
opposite chirality. In $1+1$ dimensions (one space and one time
dimension) there is no further evolution
and one can view the process as spontaneous symmetry breaking leading
to stable, non-propagating domain walls. In more than one 
spatial dimension, the
homochiralization process progresses further. If no advection
is present, only diffusion can drive the homochiralization process. 
Analytical arguments can be utilized to show that
a bubble surrounded by a region of opposite chirality tends to
shrink. Since the equations are local, the local curvature of the front
is the deciding property in determining which way a front will move.
An example of this process is shown in \Fig{pxxx}, where
we show a $2+1$ dimensional box with regions of different chirality.
The initial state in the simulation was a racemic state with small fluctuations
so that one quickly arrives at a state where there are many left and
right handed regions in the box. Here we show the evolution of
these homochiral regions. From the figure it is easy to see that the
curvature of the interface indeed determines the evolution of a region:
e.g.\ small bubbles shrink and later disappear.

Again, time scales play a crucial role in the global homochiralization
process. Using microphysically motivated values, we find that
the local relaxation to a homochiral state is rapid compared
to that of the propagation of fronts into racemic surroundings. 
In turn, this is a much faster process than the final stage
in global homochiralization when the regions of different chirality slowly
disappear as described above. Using values of molecular diffusion,
one actually finds that this time scale on Earth is much too long 
to explain the observed homochiral final state.

\begin{figure}[t!]
\resizebox{\hsize}{!}{\includegraphics{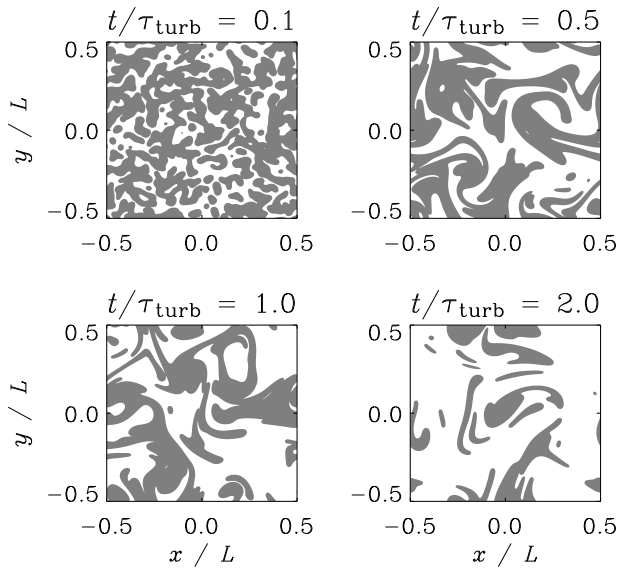}}\caption{
The spreading of homochiral regions in $2+1$ dimensions. The dark (light)
regions correspond to left (right) handed regions.
Time is given in dynamical times, $\tau_{\rm turb}=\ell/u_{\rm rms}$.
(Adapted from BM.)
}\label{turb}\end{figure}

In order to study the process further, we have also considered the
effects of advection on homochiralization. An
example in $2+1$ dimensions is shown in \Fig{turb}. The mixing
process is now greatly enhanced and one finds that the global homochiral
state is reached much more rapidly. The actual time scale depends
strongly on the strength of the flow, e.g.\ for root mean square
flow of $1$ cm/s, the time scale of global homochiralization is
of the order of 30 years. It is then clear that if the
model considered in BM captures 
the relevant features of chirality selection
on the early Earth, the effective mixing of early oceans is vital.
As the mixing is affected by many factors such as the existence
of continents and salinity of sea water, it is not difficult to 
speculate that in some secluded parts of the early oceans, life forms
of different chirality could have coexisted.

\begin{figure*}[t!]
\resizebox{\hsize}{!}{
\includegraphics[width=.33\textwidth]{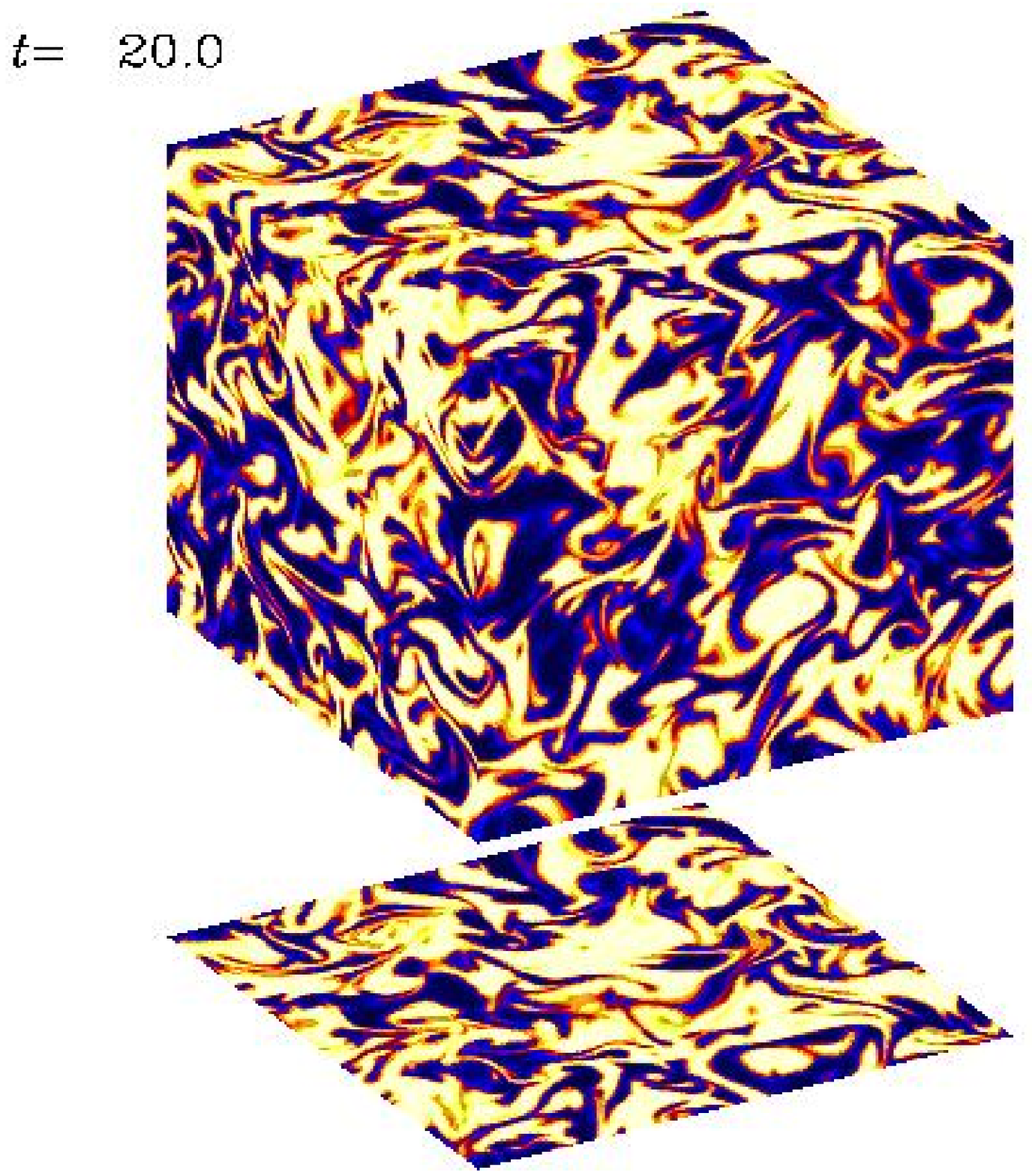}
\includegraphics[width=.33\textwidth]{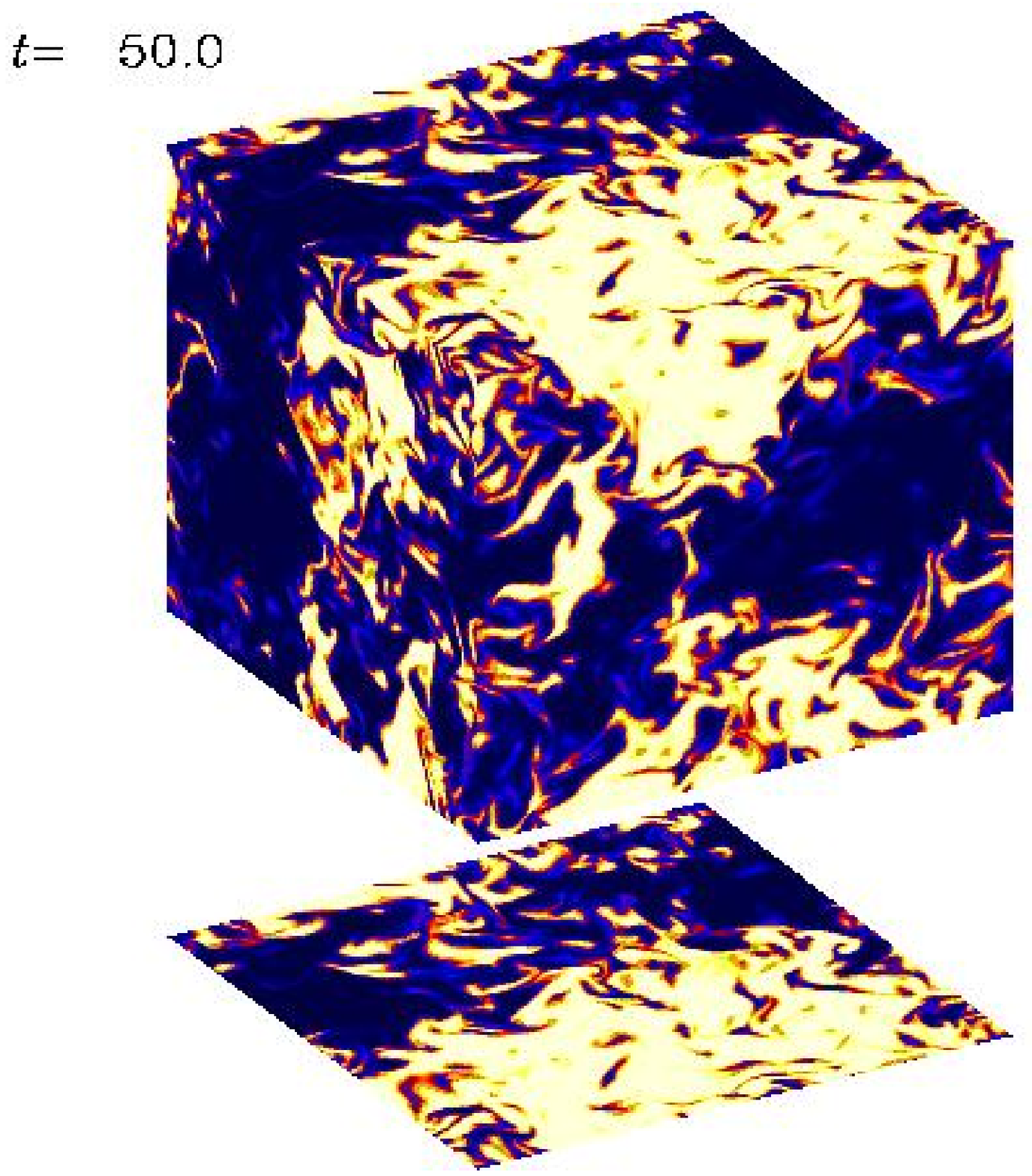}
\includegraphics[width=.33\textwidth]{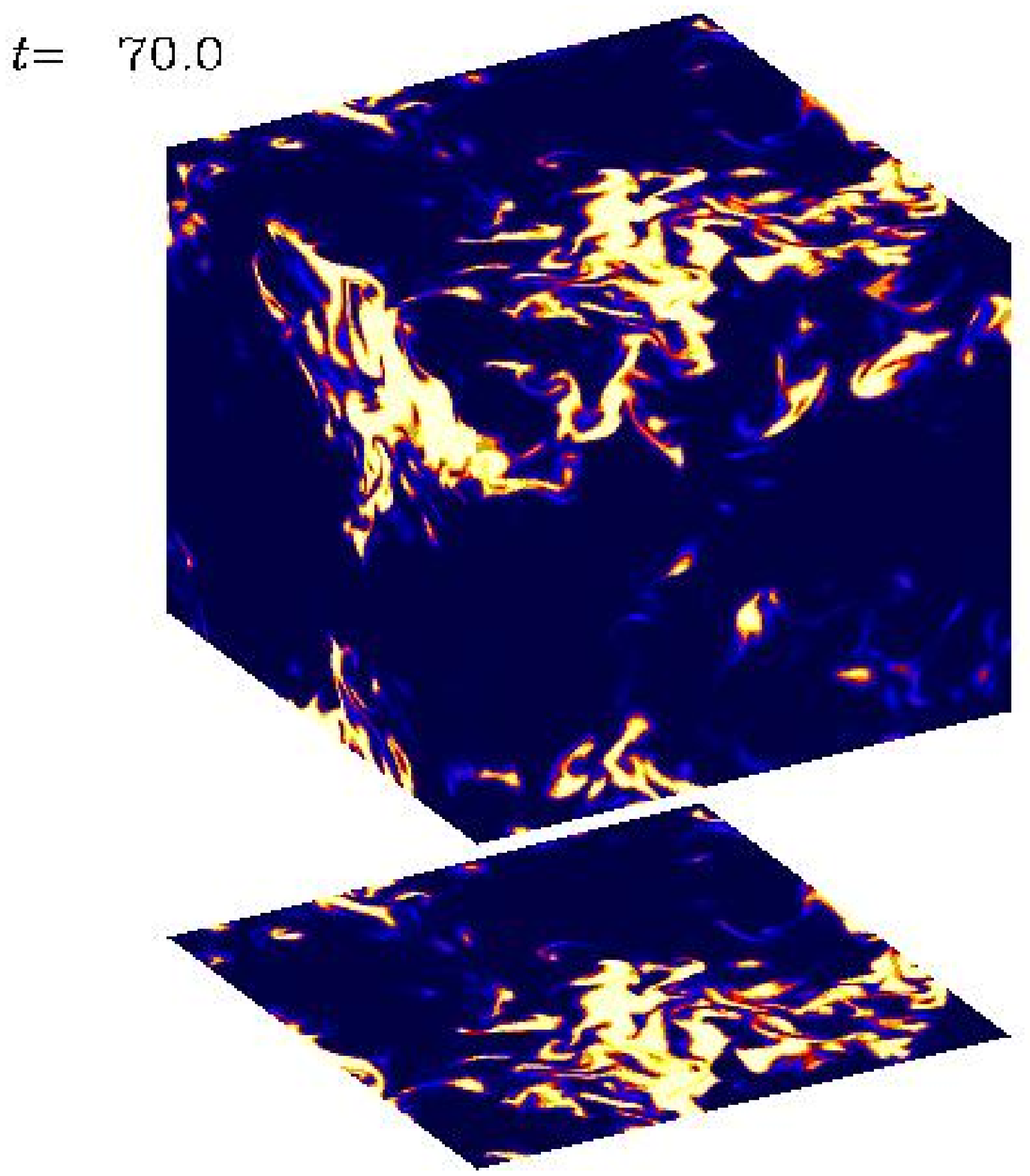}
}\caption{
The spreading of homochiral regions in $3+1$ dimensions. The dark (light)
regions correspond to left (right) handed regions.
The time $t=50$ corresponds here to $t/\tau_{\rm turb}\approx2$.
This is also the last time shown in \Fig{turb}.
By the time $t=100$ corresponds here to $t/\tau_{\rm turb}\approx4$,
the right handed life form went extinct.
}\label{img_0099}\end{figure*}

So far we have considered the spreading in two horizontal directions,
which is relevant to the Earth's surface where the vertical dynamics
may be eliminated by vertical averaging.
However, BM also considered
three-dimensional models and found that the speed of homochiralization
is somewhat enhanced (by about 30\%) relative to the two-dimensional case.
Furthermore, in both two and three dimensions
they found that the generalized enantiomeric excess,
\EQ
\eta={\bra{E_R}-\bra{E_L}\over\bra{E_R}+\bra{E_L}},
\EN
where angular brackets denote averages in space,
grows approximately linearly in time:
\EQ
{\dd\eta\over\dd t}=(12...18)\times{\cal N}
\left({2\lambda^2\over Qk_S}\right)^{1/4}{\kappa\over L^2},
\EN
where ${\cal N}$ is the number of topologically disconnected domains,
$\lambda/\sqrt{Qk_S}$ is the non-dimensional growth rate
in the spatially uniform case, $L$ is the spatial extent of the domain,
and $\kappa$ is the diffusion coefficient.

In the turbulent case, $\kappa$ can simply be replaced by a
turbulent coefficient $\kappa_{\rm turb}$, for which
BM found from their simulation the rough
estimate $\kappa_{\rm turb}\approx(0.08...0.15)u_{\rm rms}\ell$, where
$u_{\rm rms}$ is the root mean square velocity of the turbulence and
$\ell$ is the typical length scale of the turbulent eddies.
It should be emphasized, however, that such a description of turbulent
mixing is rather rudimentary and ignores, for example, the fact that
the shape of left and right handed domains is quite different
with and without turbulence:
it is more round in the case of molecular diffusion and more elongated
in the turbulent case.
The latter is a feature of the turbulence to mix fields by stretching
and folding. 
A snapshot of a fully three-dimensional simulation\footnote{Animations
of both turbulent and purely diffusive solutions can be found at
\url{http://www.nordita.dk/~brandenb/movies/chiral}} showing the
concentration of left handed building blocks is presented in
\Fig{img_0099}.

\section*{Emergence and spreading of life on the early Earth}

We have made a number of underlying assumptions in studying the
spreading of homochiral domains on the early Earth. Firstly, we
have assumed that the initial state is globally racemic.
This implies that the conditions were life could have 
emerged are uniform and favorable everywhere and,
furthermore, that a homochiral region can
spread and fill the whole domain. Secondly, we have assumed that 
the emergence of life is a common event and that it is possible
everywhere on earth, and that there are no 
mass extinctions. 
These assumptions may not necessarily 
be realistic, and one should consider what new effects can possibly 
arise from relaxing these assumptions.

Considering the possibility of mass extinctions and
that the emergence of life can be a rare event, it is clear
that a new time scale, $\tau_{\rm ext}$, is inserted into the process. 
Assuming that mass extinctions are local events (i.e.\ not all
life is destroyed in which case the process can start again),
the relative lengths of the different time scales determine the
qualitative dynamics of the system. If 
$\tau_{\rm global}\ll\tau_{\rm life}\ll\tau_{\rm ext}$, life appears
somewhere and takes over the whole system before life of opposite
chirality can emerge. Any later mass extinction events will be
insignificant as the dominant chirality will quickly win over the
achiral region. If  $\tau_{\rm life}\ll\tau_{\rm global}\ll\tau_{\rm ext}$, we
have the situation considered earlier in this paper, where
regions of different chirality compete.
If mass extinctions are common,
$\tau_{\rm ext}\ll(\tau_{\rm life},\tau_{\rm global})$, and so any emergent life
will quickly be wiped out -- at least locally.
Again, if $\tau_{\rm global}\ll\tau_{\rm life}$, it is unlikely that
life will re-emerge spontaneously in the affected areas, and these areas
will more likely be re-populated by the spreading of the homochiral
regions surrounding the now racemic area. Thus, in this case
homochirality is preserved.
On the other hand, if $\tau_{\rm life}\ll\tau_{\rm global}$, the
possibility of mass extinctions allows for new life forms to emerge
which, in turn, may prolong the time during which life forms of opposite
chirality can have coexisted.

It is not immediately obvious 
how mass extinctions should
be modeled within the framework of the BAHN model.
It seems plausible that a mass extinction due to an impact, for example,
can be modeled by removing all existing polymers.
The re-emergence of life depends on the source
term $Q$ of the substrate and on the $k_C$ parameter that controls the
rate at which new monomers are created from the substrate.
By locally suppressing $Q$ or $k_C$ as a consequence of the mass extinction,
one can model the reduced habitability by slowing down the 
generation either of the substrate or of
new monomers from the source. 
Alternatively, one could decrease the polymerization efficiency
by decreasing $k_S$ and $k_I$ on the grounds that these parameters
are likely to depend on external factors such as temperature,
salinity, and acidity.

Allowing for the additional dependencies discussed above introduces
certainly a lot of additional uncertainties, but it also removes some
of the sensitivity to initial conditions.
If the conditions
for chirality (or life) are not satisfied everywhere, the spreading
of chirality will stop and mass extinctions can eradicate
the emergent life forms. Hence, there are various approaches
to modeling the emergence of life and mass extinctions in 
the BAHN model. Either one can deplete the source of the substrate,
or one can diminish the conversion rates for monomer formation
and/or for polymerization, in which case the substrate might still
be present everywhere.

Clearly, these considerations are relevant in a more realistic model
and, furthermore, can bring interesting new dynamics into the system.
As such they are worth considering in more detail and future work
will hopefully bring us answers to these questions.
Detailed numerical models of this process may elucidate further
the possible outcomes.
We plan to adapt the {\sc Pencil Code}\footnote{
\url{http://www.nordita.dk/software/pencil-code}}, which was also
used in our earlier paper, to include the effects of a dynamical
evolution of the rate at which new left and right handed monomers
can be regenerated.

\section*{Conclusions}

We have studied the spatial evolution of chirality in an initially
racemic system. By utilizing analytical and numerical methods,
we have shown how a globally homochiral state can be reached.
The associated time scale depends strongly on the details of the system,
but molecular diffusion is too slow to homochiralize the early Earth
and turbulent flows are necessary. Effective mixing of the early oceans
drastically reduces the required time and is hence a vital
ingredient in understanding the homochiralization process.

We also consider qualitatively how a more infrequent emergence
of homochiral regions and the addition of mass extinctions
would affect the evolution of the system. Adding such (possibly more
realistic) processes enriches the dynamics and offers new interesting
avenues of research to be explored in the future.

\section*{Acknowledgments}

The Danish Center for Scientific Computing is acknowledged
for granting time on the Linux cluster in Odense (Horseshoe).

\section*{References}

\begin{list}{}{\leftmargin 3em \itemindent -3em\listparindent \itemindent
\itemsep 0pt \parsep 1pt}\item[]

Avetisov, V. A., Goldanskii, V. I., and Kuz'min,
V. V.\yjour{1991}{Phys. Today}{44}{33}
{41}{Handedness, origin of life and evolution}

Avetisov, V. A. and Goldanskii, V.\yphl{1993}{A 172}{407}
{410}{Chirality and the equation of `biological big bang'}

Bada, J. L., Luyendyk, B. P., and Maynard, J. B.\ysci{1970}{170}{730}
{732}{Marine sediments: dating by the racemization of amino acids}

Brandenburg, A., Andersen, A. C., H\"ofner, S., and Nilsson, M.\poleb{2005}
{Homochiral growth through enantiomeric cross-inhibition}
Preprints available online at: \url{http://arXiv.org/abs/q-bio/0401036} (BAHN).

Brandenburg, A. and Multam\"aki, T. (2004)
How long can left and right handed life forms coexist?,
{\it Int.\ J.\ Astrobiol.} {\bf 3}, 209-219.

Davies, P. C. W. and Lineweaver, C. H.\yab{2005}{5}{154}
{163}{Finding a second sample of life on Earth}

Frank, F. C.\yjour{1953}{Biochim.\ Biophys.\ Acta}{11}{459}
{464}{On Spontaneous Asymmetric Synthesis}

Gilbert, W.\ynat{1986}{319}{618}
{618}{Origin of life -- the RNA world}

Goldanskii, V. I. and Kuzmin, V. V.\yjour{1989}{Sov.\ Phys.\ Uspekhi}{32}{1}
{29}{Spontaneous breaking of mirror symmetry in nature and origin of life}

Hare, P. E. and Mitterer, R. M.\yjour{1967}
{Yearbook Carnegie Institution of Washington}{65}{362}
{364}{Nonprotein amino acids in fossil shells}

Joyce, G. F., Visser, G. M., van Boeckel, C. A. A., van Boom, J. H.,
Orgel, L. E., and Westrenen, J.\ynat{1984}{310}{602}
{603}{Chiral selection in poly(C)-directed synthesis of oligo(G)}

Joyce, G. F.\yjour{1991}{New Biol.}{3}{399}
{407}{The rise and fall of the RNA world}

Kozlov, I. A., Pitsch, S., and Orgel, L. E.\ypnas{1998}{95}{13448}
{13452}{Oligomerization of activated D- and L-guanosine mononucleotides
on templates containing D- and L-deoxycytidylate residues}

Saito, Y. and Hyuga, H.\yjour{2004a}{J.\ Phys.\ Soc.\ Jap.}{73}{33}
{35}{Complete homochirality induced by the nonlinear autocatalysis
and recycling}

Saito, Y. and Hyuga, H.\yjour{2004b}{J.\ Phys.\ Soc.\ Jap.}{73}{1685}
{1688}{Homochirality proliferation in space}

Sandars, P. G. H.\yoleb{2003}{33}{575}
{587}{A toy model for the generation of homochirality during polymerization}

Schmidt, J. G., Nielsen, P. E., \& Orgel, L. E.\yjour{1997}
{J. Am. Chem. Soc.}{119}{1494}{1495}{Enantiomeric cross-inhibition in
the synthesis of oligonucleotides on a nonchiral template}

Toxvaerd, S.\yjour{2004}{J. Chem. Phys.}{120}{6094}
{6099}{Domain catalyzed chemical reactions: a molecular dynamics
simulation of isomerization kinetics}

Toxvaerd, S.\ppjour{2005}{Int. J. Astrobiol.}{4}
{Origin of homochirality in biological systems}

Wattis, J. A. D. and Coveney, P. V.: 2004, Symmetry-breaking in
chiral polymerisation, {\em Orig.\ Life Evol.\ Biosph.} (in press),
arXiv:physics/0402091.

Wei-Min, L.\yjour{1982}{Orig. Life}{12}{205}
{209}{Remarks on origins of biomolecular asymmetry}

\end{list}

\vfill\bigskip\noindent{\it
$ $Id: paper.tex,v 1.26 2005/05/21 06:01:25 brandenb Exp $ $}

\end{document}